\documentstyle[twocolumn,prl,aps,graphicx,floats]{revtex}
\begin{document}
\draft
\twocolumn[\hsize\textwidth\columnwidth\hsize\csname@twocolumnfalse\endcsname

\title{Theory of temperature dependence of the Fermi surface-induced
splitting of the alloy diffuse-scattering intensity peak}
\author{Igor Tsatskis\cite{pers1}$^{,}$\cite{pers2}}
\address{Department of Earth Sciences, University of Cambridge,
Downing Street, Cambridge CB2 3EQ, United Kingdom}
\date{\today}
\maketitle

\begin{abstract}
The explanation is presented for the temperature dependence of the 
fourfold intensity peak splitting found recently in diffuse scattering 
from the disordered Cu$_{3}$Au alloy. The wavevector and temperature 
dependence of the self-energy is identified as the origin of the 
observed behaviour. Two approaches for the calculation of the 
self-energy, the high-temperature expansion and the $\alpha$-expansion, 
are proposed. Applied to the Cu$_{3}$Au alloy, both methods predict 
the increase of the splitting with temperature, in agreement with 
the experimental results.
\end{abstract}

\pacs{05.50+q, 64.60.Cn, 61.66.Dk, 71.18+y}
]

\begin{figure}
\begin{center}
\includegraphics[angle=0]{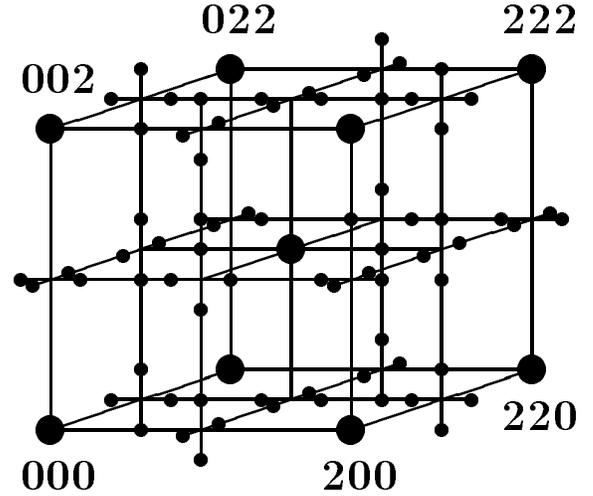}
\end{center}
\caption{Schematic reciprocal-space picture of scattering from the FCC
alloys discussed in the text. Large circles represent the Bragg
reflections, while small ones (forming characteristic crosses) correspond
to the split diffuse intensity peaks.}
\label{f1}
\end{figure}

Recently, in the first {\em in situ} experiment to resolve the fine 
structure of the equilibrium diffuse scattering intensity from the 
disordered Cu$_{3}$Au alloy, Reichert, Moss and Liang~\cite{reichert} 
have observed a marked temperature dependence of the fourfold 
splitting of the (110) short-range order (SRO) diffuse intensity 
peak. The separation of the split maxima changed reversibly, 
increasing with temperature. The same behaviour of the splitting was 
also found in~\cite{moss2} by analysing results of the Monte Carlo (MC) 
simulations for the Cu$_{0.856}$Al$_{0.144}$ alloy~\cite{roelofs}. 
The peak splitting (Fig.~\ref{f1}) is attributed to the indirect 
interaction of atoms via conduction electrons in an alloy whose Fermi 
surface has flat portions; the effective interatomic pair interaction 
itself has split minima in the reciprocal space, and their location 
is determined by the wavevector $2 {\bf k}_{F}$ spanning these flat 
portions of the Fermi surface~\cite{krivoglaz1}. As indicated 
in~\cite{reichert}, current theoretical approaches fail to explain 
the observed behaviour. Indeed, the standard approximation for the 
SRO diffuse intensity, the Krivoglaz-Clapp-Moss (KCM) 
formula~\cite{krivoglaz2}, is
\begin{equation}
I^{KCM}({\bf k}) = \frac{1}{1 + 2 c(1-c) \beta V({\bf k})} \ ,  \label{1}
\end{equation}
where $I({\bf k})$ is the intensity in Laue units, $c$ the 
concentration, $\beta=1/T$, $T$ the temperature in energy units 
and $V({\bf k})$ the Fourier transform of the combination
$V_{ij}=(V^{AA}_{ij}+V^{BB}_{ij})/2-V^{AB}_{ij}$ of potentials 
$V^{\alpha \beta}_{ij}$ with which an atom of type $\alpha$ 
at site $i$ interacts with an atom of type $\beta$ at site $j$. 
Eq.~(\ref{1}) predicts that positions of the $I({\bf k})$ peaks 
coincide with those of the corresponding minima of $V({\bf k})$; 
therefore, the splitting does not depend on $T$, if it is assumed 
that $V({\bf k})$ is $T$-independent. This assumption is justified 
at least as far as positions of the $V({\bf k})$ minima are 
concerned, since the $2 {\bf k}_{F}$ value should not change over 
the considered temperature range~\cite{reichert}. Besides, the MC 
calculations~\cite{roelofs} in which the increase of the splitting 
with temperature was found~\cite{moss2} were carried out for the 
$T$-independent pair interaction parameters. On the other hand, the 
cluster variation method~\cite{kikuchi} which in most cases leads 
to a significant improvement of the results in comparison with the 
KCM approximation~\cite{defontaine}, is practically inapplicable 
here, since interactions between atoms at distant lattice sites are 
involved (see below). 

\begin{figure}
\begin{center}
\includegraphics[angle=-90]{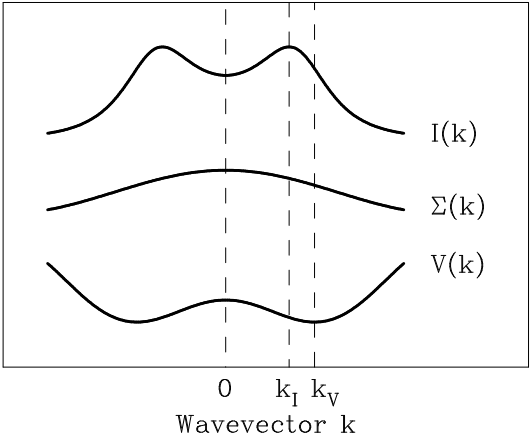}
\end{center}
\caption{Shift of the intensity peak position as a result of the 
wavevector dependence of the self-energy. The latter is as found 
for the Cu$_{3}$Au alloy (see text).}
\label{f2}
\end{figure}

The aim of the present Letter is to propose the theory of the 
temperature-dependent peak splitting observed in the Cu$_{3}$Au 
alloy. We begin by noting the exact expression for the SRO diffuse 
scattering intensity~\cite{tokar1}, 
\begin{equation}
I({\bf k}) = \frac{1}{ c(1-c) \left[ - \Sigma({\bf k}) + 
2 \beta V({\bf k}) \right] } \ ,  \label{3}
\end{equation}
where $\Sigma({\bf k})$ is the so-called self-energy which depends not 
only on ${\bf k}$, but also on $c$ and $T$. 
In the KCM approximation, however, $\Sigma$ 
is ${\bf k}$- and $T$-independent, as follows from the comparison of 
Eqs.~(\ref{1}) and (\ref{3}):
\begin{equation}
\Sigma^{KCM} = - [c(1-c)]^{-1} \, .  \label{4}
\end{equation}
Below we consider the $I({\bf k})$ profile along one of the lines 
containing split peaks, e.g., the (h10) line, and concentrate on 
two peaks around the (110) position. The peak positions $k_{I}$ 
($k$ is the deviation of the wavevector from the (110) position 
along the (h10) 
line) are determined by the condition $\partial_{k} I = 0$ which 
gives
\begin{equation}
2 \, \partial_{k} V = T \, \partial_{k} \Sigma \ .  \label{6}
\end{equation}
Eq.~(\ref{6}) means that the ${\bf k}$-dependence of $\Sigma$ leads 
to the shift $\delta k = k_{I} - k_{V}$ of the peak position with 
respect to the position $k_{V}$ of the corresponding minimum of 
$V({\bf k})$ (Fig.~\ref{f2}); $k_{V}$ is the solution of the equation 
$\partial_{k} V = 0$. Furthermore, the right side of Eq.~(\ref{6}) 
is a function of $T$, while its left side is $T$-independent. The 
$I({\bf k})$ peaks will therefore change their positions with temperature. 

At sufficiently high temperatures the behaviour of the splitting can be 
analysed by using the high-temperature expansion (HTE, in powers of 
$\beta V$) for $\Sigma$. The second-order HTE approximation 
gives~\cite{tsatskis}
\begin{mathletters}
\begin{eqnarray}
(\Sigma_{d})_{ii} & = & \Sigma^{KCM} - 
4 c (1-c) \beta^{2} \sum_{l} V_{il}^{2} \ , \\
(\Sigma_{od})_{ij} & = & 2 (1-2c)^{2} \beta^{2} V_{ij}^{2} \ .
\end{eqnarray}
\end{mathletters}
Here $\Sigma_{d}$ is the diagonal and $\Sigma_{od}$ the off-diagonal part
of~$\Sigma$. In this approximation Eq.~(\ref{6}) reduces to
\begin{equation}
\partial_{k} V = (1-2c)^{2} \beta \, \partial_{k} W \ , \ \ \ 
W_{ij}=V_{ij}^{2} \ . \label{7}
\end{equation}
The right side of Eq.~(\ref{7}) is small due to the prefactor $\beta$, 
and its solution $k_{I}$ deviates little from $k_{V}$. It is then 
sufficient to expand $\partial_{k} V$ and $\partial_{k} W$ in 
powers of the shift $\delta k$ and retain only linear terms,
\begin{mathletters}
\label{30}
\begin{eqnarray}
\partial_{k} V(k) & = & (\partial_{k}^{2} V)_{k_{V}} \, \delta k \ , 
\ \ \ (\partial_{k}^{2} V)_{k_{V}} > 0 \ , \label{16} \\
\partial_{k} W(k) & = & (\partial_{k} W)_{k_{V}}
+ (\partial_{k}^{2} W)_{k_{V}} \, \delta k \ .  \label{5}
\end{eqnarray}
\end{mathletters}
Substituting Eqs.~(\ref{30}) into Eq.~(\ref{7}) and neglecting the 
last term in Eq.~(\ref{5}) because of the smallness of the right side 
of Eq.~(\ref{7}) lead to the result
\begin{equation}
\delta k = (1-2c)^{2} \, (\partial_{k} W)_{k_{V}} \, / \, 
T \, (\partial_{k}^{2} V)_{k_{V}} \ .  \label{9}
\end{equation}
Eq.~(\ref{9}) shows two scenarios for the temperature dependence 
of the splitting, depending on the sign of $(\partial_{k} W)_{k_{V}}$. 
The first is the increase of the splitting with temperature 
discussed above. Apart from that, the theory predicts that the 
decrease of the splitting with increasing temperature is 
also possible; such temperature dependence has not yet been observed. 
The absolute value of $\delta k$ decreases as $T^{-1}$ with temperature. 
The shifts of the two peaks have opposite signs and the same 
absolute values, and the wavevector dependence of $\Sigma$ determines 
whether the splitting increases or decreases with temperature. In the 
case of an equiatomic alloy ($c=0.5$) the second-order contribution 
to $\partial_{k} \Sigma$ is zero, and the temperature behaviour of the 
shift is defined by the higher-order terms in the HTE for the self-energy.

\begin{table}
\caption{Coefficients $B_{f}$ and $C_{f}$ (Eqs.~(\ref{31})) for the
interaction $V(k)$ and related function $W(k)$ as calculated from the 
inverse MC interactions $V_{lmn}$~\protect\cite{gerold} for the Cu$_{3}$Au 
alloy at different temperatures.}
\label{t1}
\begin{tabular}{crdddd}
No. & $T$, K & $B_{V}$, meV & $C_{V}$, meV & $B_{W}$, (meV)$^{2}$ & 
$C_{W}$, (meV)$^{2}$ \\
\tableline
1 &  669 & -11.2 & 4.3 &  974.8 & 18.6 \\
2 &  678 & -11.9 & 2.8 &  202.6 &  8.1 \\
3 &  693 &  -7.5 & 3.9 &   93.8 & 15.0 \\
4 &  723 &  -3.8 &  0  &  131.6 &   0  \\
5 &  748 & -22.8 & 3.4 &  297.9 & 11.9 \\
6 &  958 &   6.7 &  0  &  -13.7 &   0  \\
7 &  958 & -71.7 & 3.1 & 1657.1 &  9.6 \\
8 & 1023 &  17.7 & 2.1 &  285.7 &  4.3 
\end{tabular}
\end{table}

We now apply the HTE to the Cu$_{3}$Au alloy for which sets of first 
8 inverse MC interactions $V_{lmn}$ were obtained at different 
temperatures using SRO parameters 
$\alpha_{lmn}$ available in the literature~\cite{gerold}. 
Despite its extended range, the behaviour of the interaction 
along the (h10) line is simple; the Fourier transform $f({\bf k})$ 
of an arbitrary FCC matrix $f_{ij}$ with non-zero elements for the 
first 20 coordination shells $lmn$ has along this line the form
\begin{equation}
f(k) = A_{f} + 2 B_{f} \cos 2 \pi k + 2 C_{f} \cos 4 \pi k \ ,  \label{11}
\end{equation}
where the relevant coefficients $B_{f}$ and $C_{f}$ are
\begin{mathletters}
\label{31}
\begin{eqnarray}
B_{f} & = & f_{200} - 4 f_{211} + 4 f_{220} + 4 f_{222} 
- 8 f_{321} \nonumber \\
& & + 4 f_{420} - 4 f_{332} + 8 f_{422} - 8 f_{521} 
+ 4 f_{442} \ ,  \label{19} \\
C_{f} & = & f_{400} - 4 f_{411} + 4 f_{420} + 4 f_{422} 
- 8 f_{431} \nonumber \\ 
& & + 4 f_{440} - 4 f_{433} + 8 f_{442} \ ,  \label{20}
\end{eqnarray}
\end{mathletters}
and $k$ is measured in the reciprocal lattice units (r.l.u.). These 
coefficients for functions $V(k)$ and $W(k)$ are shown in Table~\ref{t1}.
All four quantities scatter widely, which is the result of relatively 
low accuracy of the inverse MC interactions for the Cu$_{3}$Au alloy 
discussed in~\cite{gerold}. In particular, the accuracy and/or number 
of interactions are insufficient for the correct description of the 
split minimum of $V(k)$. The splitting in $V(k)$ occurs when $C_{V}>0$ 
and $|B_{V}|<4C_{V}$, and this is so only for sets 1 and 3. In cases 
4 and 6 $C_{V}=0$, since the eighth interaction $V_{400}$ (the only one 
contributing to $C_{V}$) was not included in the corresponding sets. In 
addition, for cases 6 and 8 $B_{V}>0$, so that the split minimum would 
occur around the $(\frac{1}{2}10)$ rather than $(110)$ position. 
Nevertheless, despite low accuracy it is seen that $B_{W}$ is positive 
and $C_{W}$ is non-negative. The only exception is set 6, where $B_{W}$ 
acquires very small negative value. However, the second set of 
interactions (set 7, with non-zero $V_{400}$) obtained using the same 
SRO parameters leads to positive values of both $B_{W}$ and $C_{W}$. 
In the case $B_{W}>0$, $C_{W} \geq 0$ function $W(k)$ has a maximum 
at the $(110)$ position which is much wider than the magnitude of 
the peak splitting; $|k_{I}|$ values observed in~\cite{reichert} 
were quite small (less than 0.1 r.l.u.). As a result, the derivative 
$\partial_{k} W$ is positive for the left and negative for the right 
minimum of $V(k)$ (Fig.~\ref{f2}), and at any finite 
temperature the intensity peaks are shifted towards the $(110)$ 
position. The absolute value of the shift increases with decreasing 
temperature, so that the splitting increases with temperature.

\begin{table}
\caption{AE coefficients $B_{f}$ and $C_{f}$ (Eqs.~(\ref{31})) for the 
self-energy $\Sigma(k)$ as calculated within the 10-shell approximation 
(except for set 16; see text) from the experimental sets of the SRO 
parameters for the Cu$_{3}$Au alloy at different temperatures.}
\label{t2}
\begin{tabular}{crccc}
No. & $T$, K & $B_{\Sigma}$ & $C_{\Sigma}$ & Ref. \\
\tableline
 1 &  669 & 0.1742 & 0.0079 & \protect\cite{chen} \\
 2 &  693 & 0.0905 & 0.0119 & \protect\cite{chen} \\
 3 &  748 & 0.0915 & 0.0013 & \protect\cite{chen} \\
 4 &  958 & 0.0268 & 0.0022 & \protect\cite{chen} \\
 5 & 1023 & 0.0231 & 0.0002 & \protect\cite{chen} \\
 6 &  669 & 0.2172 & 0.0194 & \protect\cite{bardhan} \\
 7 &  693 & 0.0787 & 0.0141 & \protect\cite{bardhan} \\
 8 &  748 & 0.0592 & 0.0035 & \protect\cite{bardhan} \\
 9 &  958 & 0.0130 & 0.0042 & \protect\cite{bardhan} \\
10 & 1023 & 0.0120 & 0.0006 & \protect\cite{bardhan} \\
11 &  678 & 0.2086 & 0.0060 & \protect\cite{cowley} \\
12 &  733 & 0.1372 & 0.0004 & \protect\cite{cowley} \\
13 &  823 & 0.0530 & 0.0006 & \protect\cite{cowley} \\
14 &  678 & 0.4137 & 0.0144 & \protect\cite{moss3} \\
15 &  723 & 0.2053 & 0.0037 & \protect\cite{moss3} \\
16 &  678 & 0.1791 &   0    & \protect\cite{walker} \\
17 &  703 & 0.0911 & 0.0021 & \protect\cite{butler} \\
\end{tabular}
\end{table}

The applicability of the HTE, similarly to that of the KCM approximation, 
is limited to the case of sufficiently high temperatures. To deal with 
moderate temperatures, we introduce here another approach which leads 
to the ${\bf k}$- and $T$-dependence of $\Sigma$, its expansion in 
powers of SRO parameters $\alpha_{ij}$ (hereafter the $\alpha$-expansion, 
or AE). Two non-zero orders of the AE for $\Sigma_{od}$ were 
calculated~\cite{tokar1} in the framework of the $\gamma$-expansion 
method (GEM)~\cite{tokar1,tokar2}:
\begin{mathletters}
\label{14}
\begin{eqnarray}
(\Sigma_{od})_{ij} & = & a \alpha_{ij}^{2} + b \alpha_{ij}^{3} + 
O(\alpha^{4}) \ , \\
a & = & \frac{(1-2c)^{2}}{2[c(1-c)]^{2}} \ , \\
b & = & \frac{[1-6c(1-c)]^{2}-3(1-2c)^{4}}{6[c(1-c)]^{3}} \ . 
\end{eqnarray}
\end{mathletters}
The expression for $\Sigma_{d}$ then comes from the sum rule 
\begin{equation}
\alpha_{ii} = \Omega^{-1} \int d {\bf k} \, I({\bf k}) = 1  \label{18}
\end{equation}
(here the integration is carried out over the Brillouin zone of volume 
$\Omega$), Eq.~(\ref{18}) being one of the AE (or GEM) equations:
\begin{eqnarray}
(\Sigma_{d})_{ii} & = & \Sigma^{KCM} + 
2 \beta \sum_{j} V_{ij} \alpha_{ij} \nonumber \\
& & - \sum_{j(\neq i)} \left( a \alpha_{ij}^{3} + 
b \alpha_{ij}^{4} \right) + O(\alpha^{5}) \ .  \label{10}
\end{eqnarray}
Note that the sum of the first two terms corresponds to the spherical 
model (SM) for SRO~\cite{hoffman}, which is the zero-order approximation 
for the AE and GEM; in the SM the self-energy is diagonal ($a=b=0$). 
The difference between the AE and GEM lies in the choice of the expansion 
parameter(s). The GEM parameter is $\gamma=\exp (- 1 / \xi)$, $\xi$ being 
the dimensionless correlation length, and terms in the diagrammatic 
expansion for the self-energy are selected according to the total length 
of all lines in the diagrams, where the line connecting sites $i$ and $j$ 
represents $\alpha_{ij}$. The GEM is based on the assumption that the 
correlations decrease rapidly with distance; this assumption is invalid 
here because distant interactions are essential. The AE uses $\alpha_{ij}$ 
themselves as the expansion parameters; the terms are chosen according to 
the number of lines in the diagrams (i.e., the powers of $\alpha_{ij}$), 
since all $\alpha_{ij}$ are sufficiently small. The GEM was successfully 
applied to both the direct and inverse problems of alloy diffuse 
scattering~\cite{tokar1,reinhard}, leading to reliable results everywhere 
except in the vicinity of the instability point. Based on our experience 
with GEM, we can as well expect the AE to be quite accurate at almost all 
temperatures.

\begin{figure}
\begin{center}
\includegraphics[angle=-90]{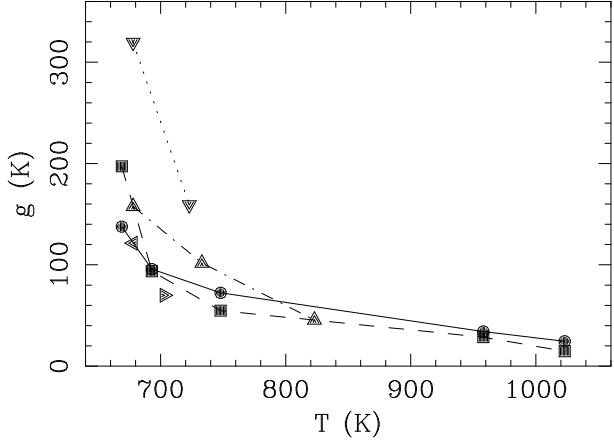}
\end{center}
\caption{Values of $g$ (Eq.~(\ref{50})) vs. temperature calculated 
using the data from Table~\ref{t2}: 
\protect\cite{chen} - dots, solid line; 
\protect\cite{bardhan} - squares, dashed line; 
\protect\cite{cowley} - upward triangles, dot-dashed line; 
\protect\cite{moss3} - downward triangles, dotted line; 
\protect\cite{walker} - leftward triangle; 
\protect\cite{butler} - rightward triangle.}
\label{f3}
\end{figure}

Applying the AE to the Cu$_{3}$Au alloy, we calculate coefficients 
$B_{\Sigma}$ and $C_{\Sigma}$ combining Eqs.~(\ref{31}) and 
(\ref{14}) and using available sets of experimental SRO 
parameters~\cite{chen,bardhan,cowley,moss3,walker,butler}. Their 
values for the case of first 10 shells included in the AE 
approximation~(\ref{14}) for the self-energy are given in Table~\ref{t2} 
(5-shell AE approximation was used for set 16, since only 5 SRO parameters 
were determined in~\cite{walker}). Inclusion of additional shells does 
not alter the results. In all cases $B_{\Sigma}$ is positive 
and $C_{\Sigma}$ is non-negative, so that, as before, $\Sigma(k)$ has a 
maximum at the $(110)$ position which is very wide in comparison with 
the peak splitting, and the intensity peaks are shifted 
towards this position. Contrary to the case of the HTE, the explicit 
temperature dependence of the AE self-energy is unknown, since the SRO 
parameters in Eqs.~(\ref{14}) are complicated functions of temperature. 
To find the temperature behaviour of the splitting, we use the data from 
Table~\ref{t2} and plot in Fig.~\ref{f3} against temperature a quantity
\begin{equation}
g = T \left| \partial^{2}_{k} \Sigma \right|_{k=0} / 8 \pi^{2} = 
T \left( B_{\Sigma} + 4 C_{\Sigma} \right) \ , \label{50}
\end{equation}
which characterises the temperature dependence of the right side of 
Eq.~(\ref{6}) at small $k$. It is seen that $g$ is a decreasing function of 
$T$, which corresponds to the increase of the splitting with temperature.
Its temperature dependence is particularly strong in the range below 800~K, 
where the intensity profile was measured in~\cite{reichert}.

To summarize, we have presented the explanation of the temperature 
dependence of the Fermi surface-induced 
diffuse intensity peak splitting found recently for the Cu$_{3}$Au 
alloy. The wavevector and temperature dependence of the self-energy 
is understood to be the origin of this behaviour. The proposed theory 
is able to describe the observed increase of the peak separation 
with temperature; in addition, it also predicts the possibility for the 
splitting to decrease as temperature increases, the behaviour 
which has not yet been found. Two methods for the calculation of the 
${\bf k}$- and $T$-dependent self-energy, the HTE and the AE, have 
been proposed. Applied to the existing experimental and inverse MC data 
for the Cu$_{3}$Au alloy, both methods predict 
the increase of the splitting with temperature, in agreement with the 
experimental findings. However, the HTE is not expected to be reliable 
when applied to alloys at realistic temperatures, so that the AE 
approach is preferable. Despite the seeming complexity of the problem 
(the interaction involves many coordination shells), the theoretical 
analysis proves to be surprisingly simple.

The author is grateful to S.C. Moss and H. Reichert for communicating 
their experimental results prior to publication and stimulating 
discussions.

\end{document}